\documentclass[reprint,twocolumn,aps,prd,superscriptaddress,floatfix,amssymb,nofootinbib]{revtex4-1}
\usepackage{verbatim}
\usepackage{mathtools}
\usepackage{bm}      
\usepackage{amssymb}  
\usepackage{natbib}
\usepackage{booktabs}
\usepackage{graphicx}
\usepackage{amsmath,amssymb}
\usepackage{import}

\newcommand{\aap}{{A\&A}}
\newcommand{\apjs}{{Astrophys.~J.~Supp.}}
\newcommand{\aj}{{Astro.~Journal}}

\newcommand{\mnras}{{Mon.~Not.~R.~Astron.~Soc.}}

\newcommand{\apjl}{{Astrophys.~J.~Lett.}}
\newcommand{\pasj}{{Publ Astron Soc Jpn }}

\begin{document}

\title{Testing emergent gravity with mass densities of galaxy clusters.}

\author{Vitali Halenka} 
\email{vithal@umich.edu}
\affiliation{Department of Physics, University of Michigan, Ann Arbor, Michigan 48109 USA}
\author{Christopher J. Miller}
\affiliation{Department of Physics, University of Michigan, Ann Arbor, Michigan 48109 USA}
\affiliation{Department of Astronomy, University of Michigan, Ann Arbor, Michigan 48109, USA}

\date{\today}

\begin{abstract}
    We use a sample of 23 galaxy clusters to test the predictions of emergent gravity (EG) \citep{Verlinde:2016toy} as alternative to dark matter. Our sample has both weak-lensing inferred total mass profiles as well as x-ray inferred baryonic gas mass profiles. Using nominal assumptions about the weak-lensing and x-ray mass profiles, we find that the EG predictions (based on no dark matter) are acceptable fits only near the virial radius. In the cores and in the outskirts, the mass profile shape differences allow us to confirm previous results that the EG model can be ruled out at $>5\sigma$. However, when we account for systematic uncertainties in the observed weak-lensing and x-ray profiles, we find good agreement for the EG predictions. For instance, if the weak-lensing total mass profiles are shallow in the core and the x-ray gas density profiles are steep in the outskirts, EG can predict the observed dark matter profile in $0.3 \le r \le 1$R$_{200}$, where R$_{200}$ is the radius which encloses 200$\times$ the critical density of the Universe. The required x-ray and lensing shapes are within the current observational systematics-limited errors on cluster profiles. We also show that EG itself allows flexibility in its predictions, which can allow for good agreement between the observations and the predictions. We conclude that we cannot formally rule our EG as an alternative to dark matter on the cluster scale and that we require better constraints on the weak-lensing and gas mass profile shapes in the region $0.3 \le r \le 1$R$_{200}$.
\end{abstract}

\maketitle

\section{Introduction}
Galaxy clusters provide a unique opportunity to study gravity in the weak-field regime. They are the only astrophysical objects which provide three simultaneous measures of gravity. We can observe the dynamical properties of clusters through the line-of-sight movement of their member galaxies. We can measure their gas content via the Bremsstrahlung x-ray emission. We can observe the distortion of spacetime through the shearing of the shapes of background galaxies. In turn, each of these needs to produce a consistent picture of the underlying gravitational theory. Our standard cosmological paradigm is based on general relativity (GR) in a de Sitter spacetime with a positive cosmological constant, where the majority of the gravitating mass is in a dark form  \citep{Frieman:2008sn}. Clusters should be able to test this theory on a case-by-case basis.

This paper is concerned with one of the biggest mysteries in modern cosmology: the origin of the dark matter, which was introduced to explain the deviation from Newtonian dynamics for galaxy rotation curves  \citep{Zwicky:1933gu, Rubin:1970zza}. Current particle theory favors options such as weakly interacting massive particles, neutrinos and axions \citep{Freese:2017idy}. Alternatively, modified Newtonian dynamics (MOND) has been shown to provide a phenomenological explanation of the galaxy rotation curves \citep{Milgrom:1983ca, Milgrom:2008rv, Famaey:2011kh}. 

Recently, there has been an advance in the theory of gravity as an emergent property of the universe. It was shown by \citet{Jacobson:1995ab} that general relativity is an emergent theory and it is possible to derive Einstein's equations from the concept of entropy of black holes and thermodynamic concepts such as temperature, heat and entropy. The revised emergent gravity (EG) proposal emphasizes the entropy content of space, which could be due to excitations of the vacuum state that manifest as dark energy \citep{Verlinde:2010hp, Verlinde:2016toy}. Briefly, this new EG defines the spacetime geometry as due to the quantum entanglement of structure at the microscopic level. Entropy then describes the information content of a gravitating system and its amount is reflected by the number of microscopic degrees of freedom.  In \citet{Verlinde:2010hp}, anti-de Sitter space was used to derive the surface entropic contribution around matter. In \citet{Verlinde:2016toy}, de Sitter spacetime was implemented in the theory which resulted in an assumed additional bulk volume component to the entropy. This volume contribution grows as the scale-size of a system increases. The excess entropy (over the surface component) results in a scale dependence for gravity as manifested through the elastic spacetime, which in turn mimics an apparent dark matter. This apparent dark matter is a result of the presence of baryonic matter.

Given the observational signature of the gas content as the dominant baryonic component in clusters, as well as the observational signature of the spacetime metric through lensing, galaxy clusters provide a rare opportunity to test EG's predictions. However, the current model proposed in \citet{Verlinde:2016toy} makes some important simplifying assumptions, such as that objects need to be spherically symmetrical, isolated, and dynamically ``relaxed''. In addition to that, \citet{Verlinde:2016toy} assumes that the universe is totally dominated by the dark energy and that implies that Hubble parameter $H(z)$ is a constant. Working in a small redshift regime is a good approximation to this assumption as it implies small changes to the Hubble parameter, which makes it to be approximately constant, as well as adds negligible corrections to the measurements due to the small change in the cosmological evolution. The real galaxy clusters which are used in the current work fit well into these assumptions as we do not include in our sample merging systems such as the Bullet cluster, and clusters with high redshifts.

Some progress has been done in testing the EG model using galaxy clusters. \citet{Nieuwenhuizen:2016uxv} tested emergent gravity with strong and weak lensing data of Abell $1689$ cluster (a part of our data sample) and showed that EG fits the data well only with inclusion of neutrinos. \citet{Ettori:2018tus} analyzed 13 clusters with reconstructed hydrostatic mass profiles and in $0.047-0.091$ redshift range and concluded that EG provides overall better fit in comparison with MOND, especially at $\sim R_{500}$ where emergent gravity mass prediction matches hydrostatic mass measurements.      

Our goal is to conduct a thorough analysis of all the available in the literature galaxy clusters data. We analyze $23$ clusters which cover a wide redshift range ($0.077-0.289$) in an extended radial range ($0.1R_{200}-2R_{200}$). Utilization of this number of clusters helps us to mitigate sample variance, which is a dominant systematic error unaddressed in \citet{Nieuwenhuizen:2016uxv}. In contrast to \citet{Ettori:2018tus}, where only weak lensing uncertainties were analyzed, we include in our analysis systematic uncertainties on the x-ray and weak lensing observables, including biases and additional scatter from the weak lensing inferred total mass profile shapes, biases from x-ray inferred baryon profile shapes, as well as stellar mass contributions and cosmology (via the Hubble parameter). 

Moreover, our cluster sample does not have issues that data of \citet{Nieuwenhuizen:2016uxv, Ettori:2018tus} possess: 13 clusters from \citet{Ettori:2018tus} have hydrostatic bias due to nonthermal pressure sources and cluster Abell $1689$ has discrepancy between mass estimates based on the x-ray data and on the gravitational lensing \citep{Broadhurst:2004bi} and it was shown by \citet{Sereno:2011fy} that Abell $1689$ has an orientation bias and the discrepancy could be resolved by dropping spherical symmetry assumption used in deriving weak lensing mass (as it was mentioned above, spherical symmetry is one of the key requirements of the EG model).

While the aim of this work is to utilize mass profiles, dynamical properties of galaxy clusters can be used as well. The idea of using escape velocity profiles of galaxy clusters to place constraints on cosmological parameters was introduced by \citet{Gifford:2013ufa, Stark2017} and it can be similarly applied to test the EG model. It should be noted that this approach has a significant statistical uncertainty due to the projection effects. However, \citet{2020arXiv200302733H} resolved this issue and showed that the observed suppression can be modeled by the function that only requires the number of observed galaxies in the projected phase-space.

In Sec. \ref{!theory} we introduce the theoretical framework of the EG model. Description of the observational data is presented in Sec. \ref{!data}. In Sec. \ref{!test_eg} the testing procedure is described as well as constraints of the EG model are presented. Discussion of the results and the conclusions are presented in Secs. \ref{!discussion} and \ref{!concl}.

For the observational data we assume a flat standard cosmology with $\Omega_M = 0.3$, $\Omega_{\Lambda} = 1 - \Omega_M$ and  $H_0 = 100h$ km s$^{-1}$ Mpc$^{-1}$ with $h = 0.7$. Throughout the paper we refer to the following quantities $R_{200}$ and $M_{200}$ which are the radius and the mass of the clusters at the point when the density drops to $200 \rho_{c, z}$, where $\rho_{c, z} = 3H^2 / (8 \pi G)$ is the critical density of the universe at redshift $z$ and $H^2 = H_0^2 (\Omega_\Lambda + \Omega_M (1+z)^3)$. The connection between $R_{200}$ and $M_{200}$ is by definition the following: $M_{200} = \frac{4\pi}{3} (200 \rho_{c, z}) R^3_{200}$.

\section{Theoretical framework}
\label{!theory}

In this section we present the main ideas of the EG model \citep{Verlinde:2016toy} as well as the equation that provides the connection between baryon matter distribution of the spherically symmetrical isolated non-dynamical system and the apparent dark matter. To do so we adopt the EG description presented in \citet{Tortora:2017uid}. 

While the original model is derived for an $n$-dimensional surface area\footnote{$\tilde\Sigma$ is used in order not to confuse our reader with $\Sigma$ which is the integral of the mass density along the line of sight }, we work in four dimensional spacetime and in a spherically symmetric approximation, such that the surface mass density is
\begin{equation}
    \tilde\Sigma(r)  = \frac{M(r)}{A(r)},
\end{equation}
where $A(r) = 4 \pi r^2$ and $M(r)$ is the total mass inside a radius r
\begin{equation}\label{mass_r}
    M(r) = \int_0^r 4 \pi r'^2 \rho(r')dr'.
\end{equation}

By incorporating quantum entanglement entropy in a de Sitter spacetime, \citet{Verlinde:2016toy} identified a thermal volume law contribution to the entropy of the universe  ($S_{DE}$). Heuristically, one can think of emergent gravity as modifying the law of gravity due to the displacement of $S_{DE}$ in the presence of matter. \citet{Tortora:2017uid} emphasizes the ``strain'' as the ratio of entropy from the baryonic matter in some volume compared to the entropy from the vacuum expansion of the universe:
\begin{equation}\label{strain}
    \epsilon_{DM}(r) = \frac{S_{DM}}{S_{DE}} = \frac{8 \pi G \tilde\Sigma_{DM}(r)}{a_0},
\end{equation}
where $a_0 = c H_0$ is the acceleration scale \citep{Milgrom:1983ca}.
In regions of normal matter density with a large number of microscopic states $\epsilon_{DM}(r) > 1$, the theory recovers the simple Newtonian equations as a limit to the theory of general relativity. However, as the number of microscopic states becomes small (i.e., in low density regions of the Universe) ($\epsilon_{DM}(r) < 1$), not all of the de Sitter entropy ($S_{DE}$) is displaced by matter. The remaining entropy modifies the normal gravitational laws in the GR weak-field limit (i.e., the Newtonian regime). This gravitational effect can be described by an additional surface density component,
\begin{equation}\label{surface_density}
    \tilde\Sigma_{DM} = \frac{a_0 \epsilon_{DM}}{8 \pi G},
\end{equation}
where the subscript $DM$ refers to the apparent dark matter.

To get the "mass" of the apparent DM one needs to estimate the elastic energy due to the presence of the baryonic matter.
The calculations (see \citet{Verlinde:2016toy}) lead to the following inequality
\begin{equation}\label{inequality}
    \int_{\mathcal{B}} \epsilon^2_{DM} dV \leqslant V_{M_b}(\mathcal{B}),
\end{equation}
where $\epsilon_{DM}$ is defined in formula \ref{strain} and $\mathcal{B}$ is the spherical region with the area $A(r)= 4 \pi r^2$ and radius $r$. The r.h.s. of the inequality \ref{inequality} is the volume which contains an equal amount of entropy with the average entropy density of the universe to the one which is removed by the presence of baryons,
\begin{equation}\label{baryon_volume}
    V_{M_b}(r) = \frac{8 \pi G r M_b(r)}{3 a_0},
\end{equation}
where $M_b(r)$ is the total mass of the baryonic matter inside some radius $r$.

\citet{Tortora:2017uid} notes that most of the recent papers on the EG theory focus on the equality in the expression \ref{inequality}, but there is no particular reason to choose this case as it places the upper bound on the amount of the apparent DM. However, if we work at the maximum we can combine equations \ref{surface_density} and  \ref{baryon_volume} with equality in \ref{inequality} to get:
\begin{equation}\label{Main_eg}
    M_b(r) = \frac{6}{a_0 r}\int_0^r \frac{G M_{DM}^2(r')}{r'^2} dr'.
\end{equation}
To find the apparent dark matter we can differentiate both sides of the Eq. (\ref{Main_eg})
\begin{equation}\label{DM_from_Mbaryon}
    M_{DM}(r) = \Bigr[ \frac{a_0 r^2}{6G} \Bigr(M_b(r) + r\frac{\partial M_b(r)}{\partial r} \Bigr) \Bigr]^{0.5}.
\end{equation}

Equations (\ref{Main_eg}) and (\ref{DM_from_Mbaryon}) provide predictions from the theory to test the data against. We use the observed baryonic matter density through the emitting x-ray gas combined with a total (dark matter plus baryonic matter) inferred from weak lensing to make these tests. 

\begin{table*}[t]
\caption{List of galaxy clusters and references}
\label{table1}
\begin{tabular}{@{}|l|l|l|l|l|l|l|l|l|l|l|l|@{}}
\toprule
\hline 
Cluster name\footnote{The original papers are cited above, but actual spherical weak lensing masses (and their respective errors) we use in our analysis were taken from the \citet{Sereno:2014aea} meta catalog. More specifically, \citet{Sereno:2014aea} standardizes the $M_{200}$  masses for the clusters shown above (as inferred from each reference listed in the "weak lensing" column) for the fiducial cosmology mentioned in our Introduction.} & Redshift & Weak lensing\footnote{The abbreviations in this column refer to the following papers: H15= \citet{hoekstra2015}, OK08 = \citet{okabe2008}, OK10 = \citet{Okabe2010}, OK15= \citet{okabe2015}, A14 = \citet{Applegate2014}, C04 = \citet{Cypriano2004}, D06 = \citet{Dahle:2006fa}, P07 = \citet{Pedersen2007}, U15= \citet{Umetsu2015}. We averaged over multiple weak lensing sources to get $M_{200}$ as well as the errors of the clusters A2219 and A773.} & $M_{200, w}$ & $R_{200, w}$ & $\rho_{0, w}$ & $r_{0, w}$ & $n_w$ & Baryons\footnote{The abbreviations in this column refer to the following papers: G17 = \citet{Giles:2015waa}, V06 = \citet{Vikhlinin:2005mp}, Gi17 = \citet{Giacintucci:2017xyd}} & $\rho_{0, b}$ & $r_{0, b}$ & $n_b$ \\ 
 &  & & $(10^{14} M_{\odot})$\footnote{Index $w$ stands for weak lensing in the Einasto parameters (\ref{Ein_density})} & (Mpc) & $(10^{17} M_\odot)$ & $(pc)$ & & & $(10^{15} M_\odot)$\footnote{Index $b$ stands for baryon gas in the Einasto parameters (\ref{Ein_density})}& $(pc)$ &  \\ 
\hline 
\midrule

A1682	&	0.227	&	P07	&	6.05	&	1.62	&	6.1	&	65.8	&	4.21	&	G17	&	1.62	&	8980	&	2.89	\\
A1423	&	0.214	&	OK15	&	6.7	&	1.68	&	5.8	&	71.9	&	4.19	&	G17	&	40.5	&	20.8	&	5.08	\\
A2029	&	0.077	&	C04	&	10.28	&	2.03	&	5.2	&	86.3	&	4.19	&	V06	&	54.0	&	111.6	&	4.2	\\
A2219	&	0.226	&	OK10/0K15/A14	&	15.33	&	2.21	&	4.46	&	122.7	&	4.13	&	G17	&	4.63	&	6347.8	&	2.95	\\
A520	&	0.201	&	H15	&	12.75	&	2.09	&	4.63	&	111.6	&	4.14	&	G17	&	0.46	&	97100	&	1.8	\\
A773	&	0.217	&	OK15/D06	&	15.45	&	2.22	&	4.43	&	123.7	&	4.13	&	G17	&	8.36	&	1670	&	3.36	\\
ZwCl3146	&	0.289	&	OK15	&	7.94	&	1.73	&	5.36	&	86.6	&	4.15	&	G17	&	1170.0	&	1.8	&	5.38	\\
RXJ1720	&	0.16	&	OK10	&	5.38	&	1.59	&	6.43	&	58	&	4.23	&	G17	&	250.0	&	7.1	&	5.07	\\
RXCJ1504	&	0.217	&	OK15	&	8.26	&	1.8	&	5.46	&	81.2	&	4.18	&	Gi17	&	1280.0	&	0.9	&	5.58	\\
A2111	&	0.229	&	H15	&	8.08	&	1.78	&	5.38	&	83.5	&	4.17	&	G17	&	9.49	&	535	&	3.9	\\
A611	&	0.287	&	OK10	&	8.68	&	1.78	&	5.19	&	92.2	&	4.15	&	G17	&	260.0	&	6.3	&	5.12	\\
A697	&	0.281	&	OK10	&	15.16	&	2.15	&	4.47	&	125.9	&	4.12	&	G17	&	3.16	&	11500	&	2.67	\\
A1689	&	0.184	&	U15	&	18.86	&	2.4	&	4.2	&	137.2	&	4.12	&	Gi17	&	311.0	&	3.9	&	5.29	\\
A1914	&	0.166	&	H15	&	11.2	&	2.03	&	4.89	&	99	&	4.16	&	G17	&	74.51	&	174	&	3.95	\\
A2261	&	0.224	&	OK15	&	18.01	&	2.33	&	4.25	&	135.7	&	4.12	&	G17	&	526.0	&	1.1	&	5.79	\\
A1835	&	0.251	&	H15	&	16.88	&	2.26	&	4.35	&	131.3	&	4.12	&	G17	&	568.0	&	4.9	&	5.15	\\
A267	&	0.229	&	OK15	&	9.07	&	1.85	&	5.26	&	87.7	&	4.17	&	G17	&	383.0	&	2.2	&	5.48	\\
A1763	&	0.231	&	H15	&	14.13	&	2.14	&	4.48	&	120.9	&	4.12	&	G17	&	2.19	&	11000	&	2.75	\\
A963	&	0.204	&	OK15	&	10.66	&	1.97	&	4.95	&	97.9	&	4.15	&	G17	&	2.36	&	14634	&	2.42	\\
A383	&	0.189	&	OK15	&	8.06	&	1.8	&	5.54	&	78.2	&	4.19	&	V06	&	450.0	&	1.9	&	5.39	\\
A2142	&	0.09	&	OK08	&	13.63	&	2.22	&	4.74	&	104.4	&	4.16	&	Gi17	&	333.0	&	1.1	&	5.86	\\
RXCJ2129	&	0.234	&	OK15	&	7.24	&	1.71	&	5.67	&	75.8	&	4.18	&	G17	&	23.8	&	443	&	3.73	\\
A2631	&	0.277	&	OK15	&	12.34	&	2.02	&	4.7	&	112.5	&	4.13	&	G17	&	1.11	&	36800	&	2.17	\\

\hline 
\bottomrule

\end{tabular}
\end{table*}

\section{Data}
\label{!data}
We require inferred total mass and baryonic mass profiles for a large set of galaxy clusters. The weak lensing data are given in the NFW formulism \citet{Navarro:1995iw}. The baryonic data are given via a $\beta$ profile \citet{Vikhlinin:2005mp}. Because we are going to focus on the virial region of clusters, we simplify the analysis by using a single analytical form for all of the mass profiles. There has been much recent work \citep{2006AJ....132.2685M, Miller:2016fku} on the dark matter mass profiles of clusters in simulations that show that the preferred profile is close to an Einasto form \citep{Einasto:1965czb}. A great advantage of the Einasto parametrization over the NFW or the $\beta$ form in the context of gravitational studies is that it predicts a fixed mass of a cluster, i.e. $M(r)$ (\ref{mass_r}) converges to a particular number. The Einasto profile is described by
\begin{equation}\label{Ein_density}
    \rho(r) = \rho_0 \exp(-s^{1/n}),
\end{equation}
where $s \equiv \frac{r_0}{r}$, $r_0$ is the scale radius, $\rho_0$ is the normalization and $n$ is the power index. Below, we discuss how we convert between the Einasto and the NFW or $\beta$ models, as well as the implication of this profile homogenization.

\subsection{Total mass profiles} \label{weak_lensing_data}
We are using Sereno meta catalog \citep{Sereno:2014aea} as a source of weak lensing data of the galaxy clusters. The weak lensing parameters are presented in the NFW form \citep{Navarro:1996gj}
\begin{equation}
    \rho_{NFW} = \frac{\rho_s}{ \frac{r}{r_s} (1 + \frac{r}{r_s})^2 },
\label{NFW_profile}
\end{equation}
where $\rho_s$ and $r_s$ are two parameters of the model and we can define concentration parameter $c_{200} = r_{200} / r_s$, which describes the overall shapes of the density profiles. \citet{Sereno:2014aea} uses the following relationship between $M_{200}$ and $c_{200}$
\begin{equation}\label{conc_sereno}
    c_{200} = A \Bigr(\frac{M_{200}}{M_{pivot}}\Bigr)^B (1+z)^C, 
\end{equation}
where $A = 5.71 \pm 0.12$, $B = -0.084 \pm 0.006$, $C = -0.47 \pm  0.04$, $M_{pivot} = 2 \times 10^{12} M_{\odot}/h$ \citep{2008MNRAS.390L..64D}.

We convert the NFW profiles to the Einasto form (\ref{Ein_density}). \citet{Sereno:2015ula} has already showed that both the NFW and the Einasto density profiles are nearly identical outside the core region of clusters up to $R_{200}$. We confirm this and find that the Einasto parametrization can recreate a given NFW profile in the region $0.15 \leqslant r \leqslant R_{200}$ to less than 1\% accuracy. This defines the statistical floor of our total mass profiles. We include additional error on the total mass profiles from the published errors in \citep{Sereno:2014aea}.

The use of a specific mass versus concentration relationship adds a systematic uncertainty from the observations. The average concentration of our sample is $<c_{200}> = 3.15$ and individual concentrations are in the range $2.57 < c_{200} < 3.58$. We also explore the effect of an additional systematic error in the concentrations on our conclusions.

\subsection{Baryon profiles}
In what follows we are using only the gas density profile as a source of baryon density while neglecting stellar mass contribution as it is around or less than $10\%$ of the overall baryon mass for the clusters with the masses of the clusters we use in our analysis \citep{2009ApJ...703..982G, 2010MNRAS.407..263A, 2013A&A...555A..66L}. We will test the assumption of neglecting stellar contribution later in the text. 
Also, we do not take into account the brightest cluster galaxy (BCG) in each of the galaxy clusters, since it was shown by \citet{ZuHone:2019hdt} that the BCG contribution is negligible outside $r \sim 100$ kpc (in our analysis, we focus on the region outside $r \sim 0.1 \times R_{200}$ which is $r \sim 160-240$ kpc for the analyzed clusters (see table \ref{table1})). The gas density profiles are taken from several sources \citet{Giles:2015waa, Vikhlinin:2005mp,  Giacintucci:2017xyd}. Unlike the weak lensing data, the baryon density uncertainties are not reported in the papers from which the data used in this work were extracted.

\begin{figure}[ht] 
\centering
\includegraphics[width=1.02\linewidth]{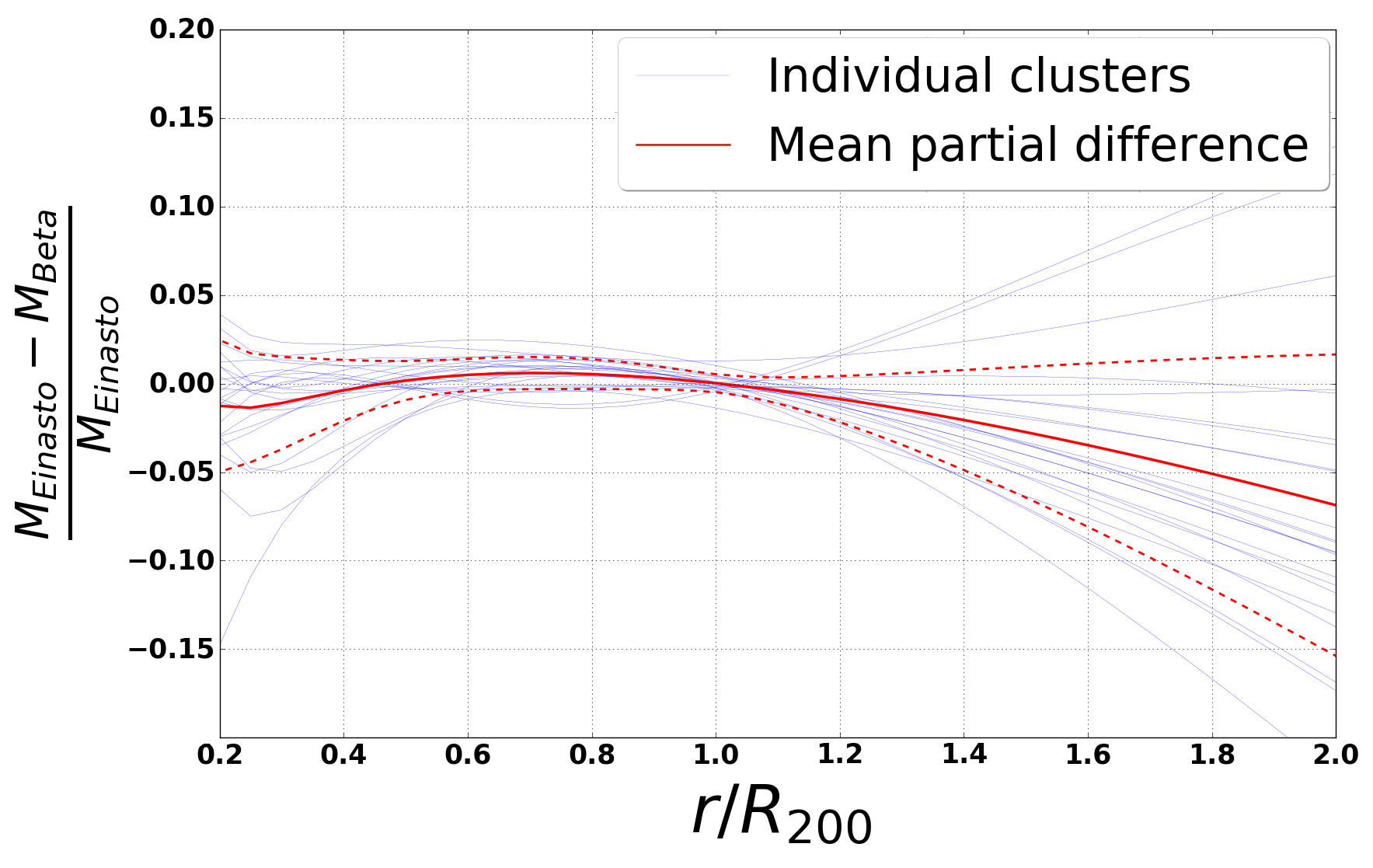}
\caption{Partial difference between Einasto and beta profiles. Blue lines are the partial differences of individual clusters. Red solid line is the mean value and dashed lines are $68.3\%$ error bars around the mean. As we can see they are almost identical all the way until $R_{200}$ and starts to deviate outside this range. Moreover, the beta profile at average tends to overestimate the mass $M(r)$ as the partial difference is smaller than zero after $R_{200}$. }
\label{Ein_beta}
\end{figure}

\citet{Giles:2015waa, Vikhlinin:2005mp} use beta profile to infer the baryon density distribution,
\begin{multline}\label{beta_profile}
    n_p n_e = n_0^2 \frac{ (r/r_c)^{-\alpha} }{(1+ r^2/r_c^2)^{3\beta - \alpha/2}} \frac{1}{(1+ r^\gamma / r_s^\gamma)^{\epsilon/\gamma}} + \\ +
    \frac{n_{02}^2}{(1+ r^2/r_c^2)^{3\beta_2}}, 
\end{multline}
where $n_p$ and $n_e$ are the number densities of protons and electrons in a gas, $r_c$ is the characteristic radius and $n_0$ is the central density. \citet{Giles:2015waa} uses the same profile but without the second term in the sum, i.e. without $\frac{n_{02}^2}{(1+ r^2/r_c^2)^{3\beta_2}}$.

To get the actual baryon matter density distribution, relation \ref{beta_profile} is used \citep{Vikhlinin:2005mp},
\begin{equation}
    \rho_b = 1.624 m_p (n_p n_e) ^ {0.5},
\end{equation}
where $m_p$ is the proton mass.

\citet{Giacintucci:2017xyd} uses so called double beta model that provides the number density of the electrons in the gas
\begin{equation}\label{double_beta}
    n_e = \frac{n_0}{1 + f} \Bigr( (1 + \frac{r^2}{r^2_{c_1}})^{-1.5 \beta_1} + f(1 + \frac{r^2}{r^2_{c_2}})^{-1.5\beta_2} \Bigr),
\end{equation}
where $n_0$ is the central density, the rest of the parameters are free parameters and in order to infer the baryon matter profile the following relation is used \citep{Schellenberger:2017wdw}
\begin{equation}
    M_b(r) = 4.576 \pi m_p \int_0^r n_e (r') r'^2 dr'.
\end{equation}

We transform the beta profiles into Einasto profiles in the identical manner as the NFW profiles (see subsection \ref{weak_lensing_data}). The Einasto profile recreates the beta profile with a high precision in the region from around the core until $R_{200}$ (see Fig. \ref{Ein_beta}). While we chose to transfer beta to the Einasto profile in the region up to $R_{200}$, we could do this procedure with almost identical accuracy in the region up to $2R_{200}$.

We note that like for the case of the weak lensing profiles, the shapes of the baryon profiles are systematics limited.  In Eq. (\ref{beta_profile}), the parameter $\epsilon$ governs the shape of the baryon profile in the outskirts. Large values indicate steeper slopes. \citet{Vikhlinin:2005mp} applies an upper limit of $\epsilon=5$ and his original sample has $\langle \epsilon \rangle = 3.24$. On the other hand, the fits to our subset of the cluster data by Eq. (\ref{beta_profile}) have significantly shallower slopes at $\langle \epsilon \rangle = 1.69$. Uncertainties on $\epsilon$ are not available, and so like concentration in weak lensing NFW fits, we explore systematic errors in this parameter later on.

\subsection{Dark matter profiles}\label{section:DM_profile}
In what follows, we treat the weak lensing masses as total masses of the galaxy clusters and the dark matter mass is calculated as
\begin{equation}\label{M_tot}
    M_{DM} = M_{tot} - M_b,
\end{equation} 
where $M_{tot}$, $M_{DM}$ and $M_b$ are the total mass, the dark matter mass and the baryon matter mass of a cluster.

\subsection{The clusters}\label{subsect - clusters}
We list all the 23 clusters in the Table \ref{table1}. The average mass of our set of 23 observed galaxy clusters is $<M> = 1.14 \times 10^{15} M_{\odot}$ while individual masses are in rather broad range $(5.4 \times 10^{14} M_{\odot}, 1.89 \times 10^{15} M_{\odot})$. To create a list of galaxy clusters used in this work, the following selection procedure was followed. The first criteria is the data availability, i.e. only clusters with the available in the literature weak lensing and baryon density profiles were selected. The second stage is to remove from the sample merging systems (e.g., the Bullet cluster) and clusters with high redshifts (e.g. BLOXJ1056 with $z=0.831$). All of the clusters have rather small redshifts ($<0.289$) and that fits well into approximation made by the EG theory, i.e. constant Hubble parameter. However, we will still test this assumption later in the current manuscript.

\section{Testing emergent gravity}\label{!test_eg}
We have two ways of comparing the EG model with the data. The first one is based on Eq. (\ref{Main_eg}) such that we compare the observed baryon mass profile to the one predicted from the ``observed'' dark matter profile. Recall from Section \ref{section:DM_profile} that the observed dark matter profile is actually the total mass profile from weak lensing minus the observed baryon profile. The second approach is based on Eq. (\ref{DM_from_Mbaryon}) that represents the opposite situation. In this case, we use the observed baryon profile to make a prediction for the dark matter profile and compare that to the ``observed'' dark matter profile. 

\subsection{Qualitative assessment of the EG model}\label{M_prediction}
\begin{figure}[t]
\centering
\includegraphics[width=1\linewidth]{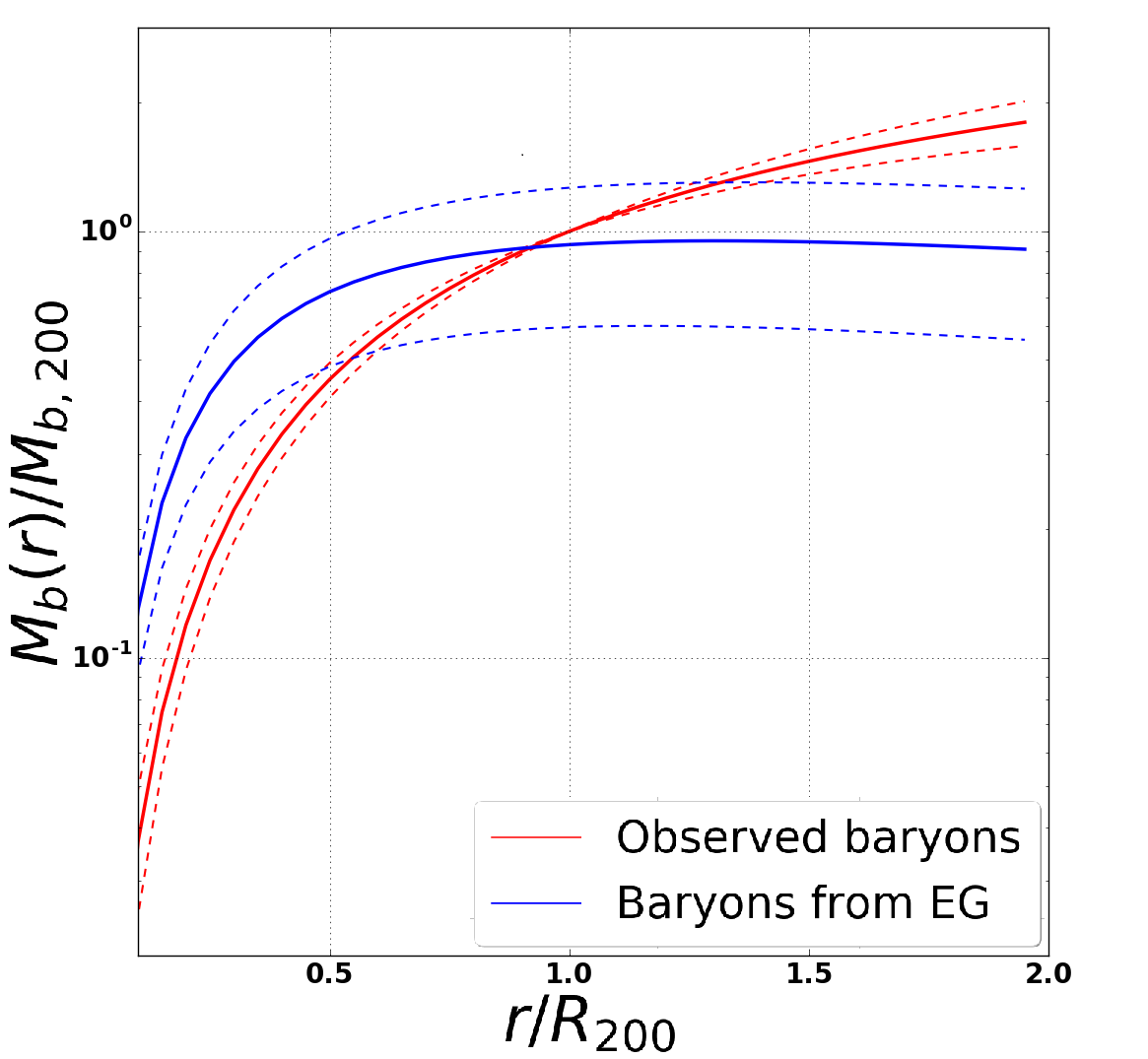}
\caption{The normalized by $M_b$ at $R_{200}$ average total baryon mass inside a spherical region of a radius r (see f-la \ref{mass_r}) for all the 23 galaxy clusters from the data of the baryon density distribution (red lines) and by applying EG relation \ref{Main_eg} to the dark matter from the data (blue lines). Solid and dashed lines are the means and $68.3\%$ error bars around the means. The baryon density here was increased by 10\% to account for the stellar mass. Note the agreement in the total baryonic mass at $\sim R_{200}$, except that EG predicts most of the baryons to be in the cluster cores. 
}
\label{plot:Mb}
\end{figure}

\begin{figure}[t] 
\centering
\includegraphics[width=1\linewidth]{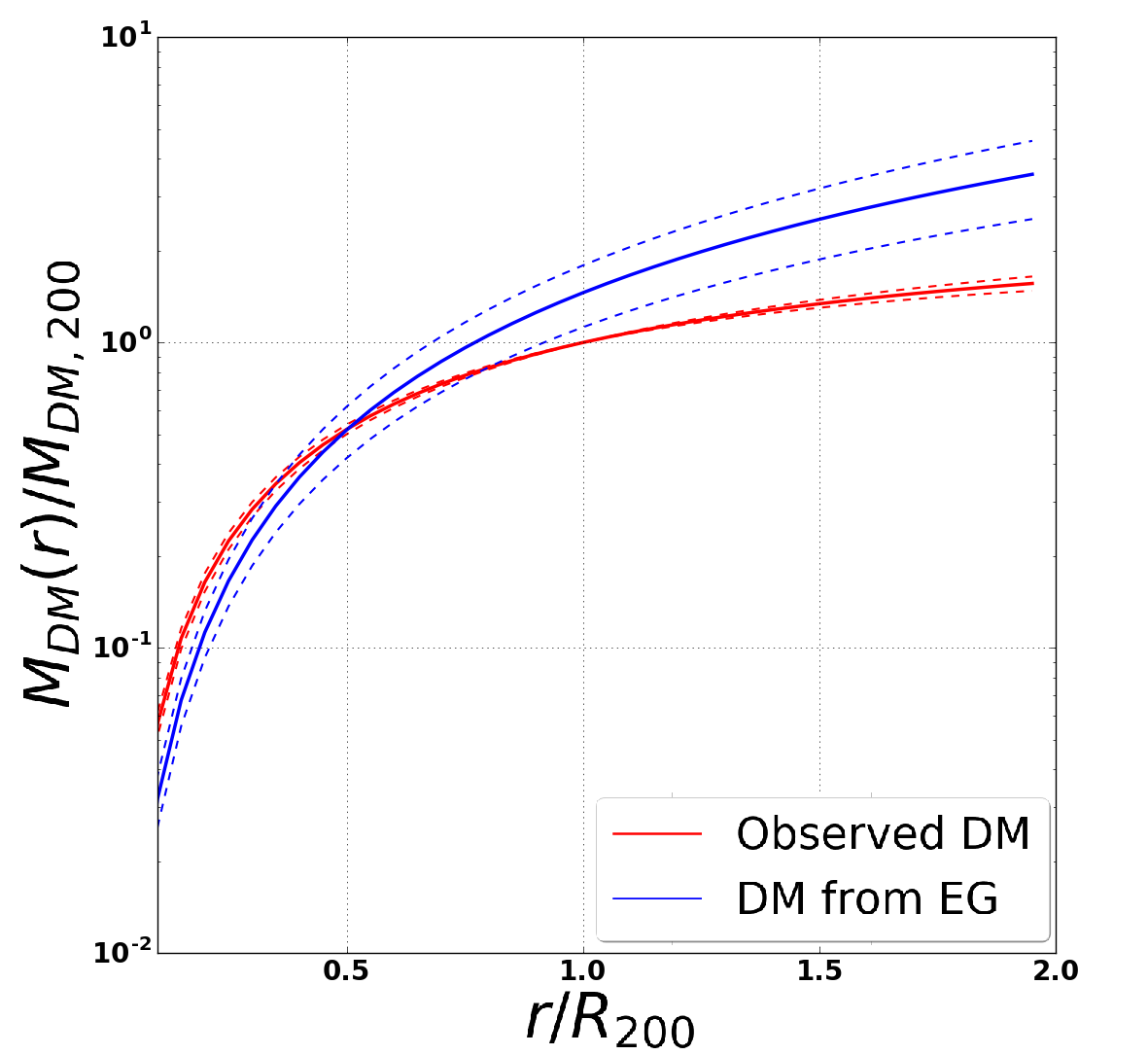}
\caption{The normalized by $M_{DM}$ at $R_{200}$ average total dark matter mass inside a spherical region of a radius r (see f-la \ref{mass_r}) from the data (red lines) and by applying EG relation \ref{DM_from_Mbaryon} to the baryon density distribution data (blue lines). Solid and dashed lines are the means and $68.3\%$ sample variance around the means. Baryon density here was increased by 10\% to account for the stellar mass. One might be able to notice that blue line increases linearly starting from around  $R_{200}$ which does not look physical as we expect the mass of the galaxy clusters to stop growing at some finite radius close to a few $R_{200}$. Moreover, we see significant difference between blue and red solid lines especially at high radii.}
\label{plot:MD}
\end{figure}

Fig. \ref{plot:Mb} shows the results of applying Eq. (\ref{Main_eg}) that makes a prediction for the baryon profile from the dark matter profile. The red line is the observed baryon profiles using the x-ray data and including a 10\% additional stellar component. The blue line comes from applying Eq. (\ref{Main_eg}) using the dark matter mass profile from Eq. (\ref{M_tot}). We note that  for clarity in Figs. \ref{plot:Mb} and \ref{plot:MD}, we normalized each cluster baryon profile by its value at $R_{200}$. Actual radii (in terms of Mpc) were used in all of the statistical analyses. The solid lines represent the means of the samples and the dashed lines the observed 1$\sigma$ scatter from the 23 systems. We find that the data (red) and the model (blue) agree at $\sim R_{200}$ and beyond.  However, EG predicts that the majority of the baryons are enclosed within the cluster core. Specifically, EG predicts that 50\% of the baryons are within $\sim 0.2\times R_{200}$. However, the observed baryons do not reach 50\% until $\sim 0.5\times R_{200}$.

Fig.  \ref{plot:MD} shows the results of applying Eq. (\ref{DM_from_Mbaryon}) that makes a prediction for the dark matter profile from observed baryon profile. The red line is from the observed dark matter profiles. The blue line comes from applying Eq. (\ref{DM_from_Mbaryon}) to the observed baryon profiles. The solid lines represent the means of the samples and the dashed lines the observed 1$\sigma$ scatter from the 23 systems. We normalize each of the cluster's dark matter profiles by its value at the weak-lensing inferred $R_{200}$ in order to conduct a combined analysis of all 23 galaxy clusters.

From Figs. \ref{plot:Mb} and \ref{plot:MD} we find a qualitative agreement between the observations and EG theory. A key success of the theory is the amplitude it predicts as it is close to what we observe near the virial radius. In other words, using just the observed baryons, EG predicts the observed dark matter mass at $\sim R_{200}$. Likewise, the difference between the total weak-lensing inferred mass and the baryon mass at $\sim R_{200}$ is predicted from EG using just the baryons alone. However, differences become apparent at smaller and larger\footnote{One can notice strange behaviour in EG predictions at high radii which is especially noticeable on Fig. \ref{plot:Mb}, where $M_b(r)$ starts to decrease at $\sim 1.5 \times R_{200}$. This result can be derived analytically: Eq. (\ref{Main_eg}) leads to $M_b(r) \propto \frac{1}{r^2}$ assuming convergence of $M_{DM}(r)$ to a constant number at high radii.} radii. Unfortunately, the observed baryon profiles are not highly constraining in the core regions and in the outskirts of clusters. The cores of clusters are active environments with varying levels of astrophysical processes which could alter the profiles. Likewise, x-ray surface brightnesses drop steeply beyond $R_{500}$, to the point where it becomes impossible to constrain the gas density profile out beyond the virial radius. We discuss these issues in the next subsections. In the meantime, we can first apply a more stringent quantitative comparison in the region where the data is more certain.

\begin{figure}[t] 
\centering
\includegraphics[width=1.03\linewidth]{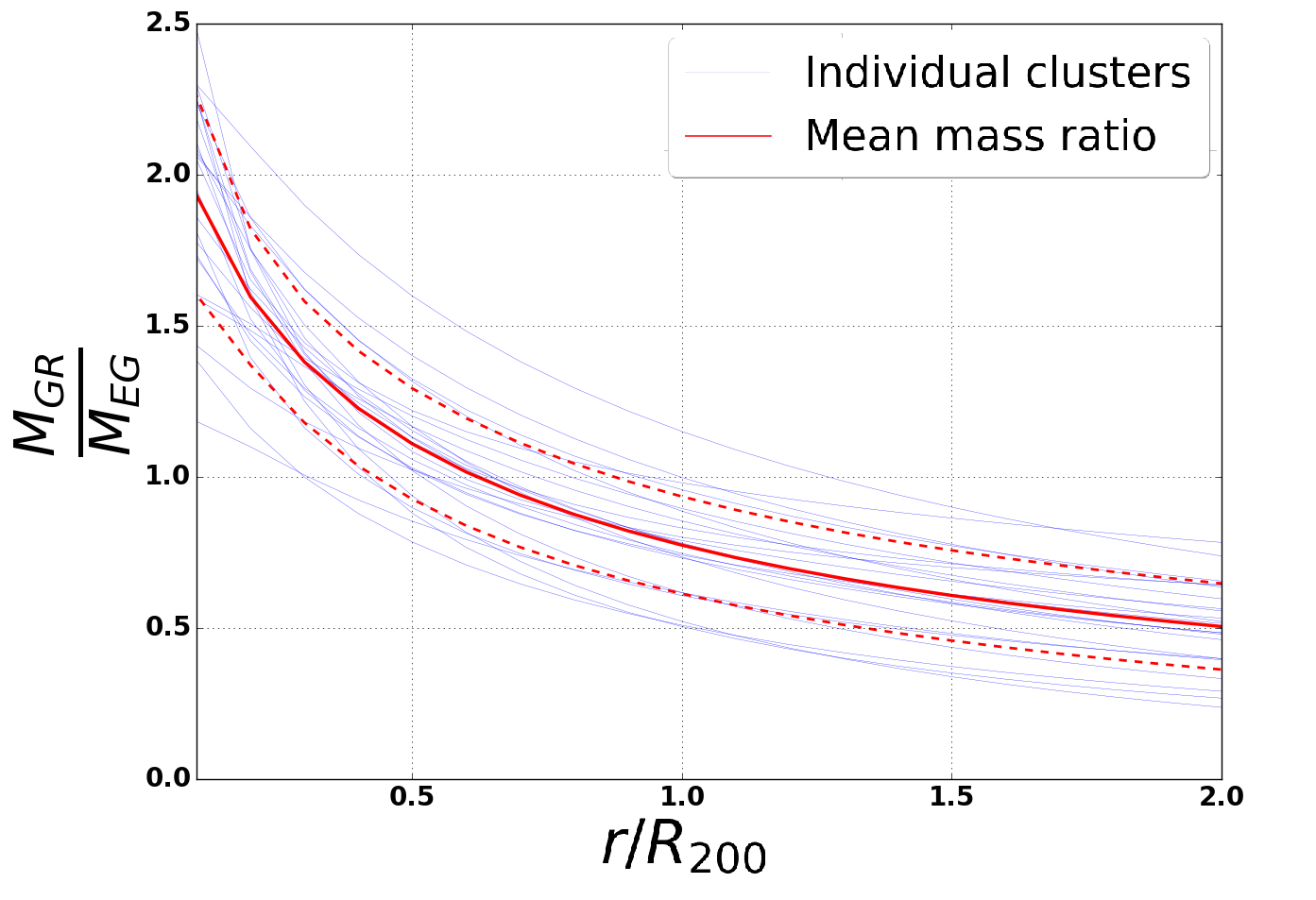}
\caption{The mass ratio $\frac{M_{GR}}{M_{EG}}$ of the observed dark matter ($M_{GR}$) to the predicted by the EG model apparent dark matter ($M_{EG}$). Thing blue lines are the individual mass ratios of the real 23 galaxy clusters. Red solid and dashed lines are the mean and $68.3\%$ error bars around the mean of all the blue lines. In order for the EG model to be compatible with the observational data the red mean line should be as close as possible to the unity. Unfortunately, this is not the case all the way until approximately $0.6R_{200}$ when the red dashed line crosses unity. This result means that the EG model does not describe the observed data in all the regions except $\sim 0.6R_{200}$, i.e. the EG model underestimates the amount of matter close to the core and overestimates the mass at high radii. }
\label{mass_ratio_23}
\end{figure}

\subsection{Data analysis and statistical constraint of the EG model}

To compare the EG model with the data we apply fitting procedure which is based on minimization of $\chi^2$
\begin{equation}\label{chi_fla}
    \chi^2 = \sum_i \frac{(M(r_i) - M_{th}(r_i))^2}{\sigma(r_i)^2 },
\end{equation}
where $M_{th}(r_i)$ is given by the r.h.s. of the Eq. (\ref{DM_from_Mbaryon}) (the apparent dark matter prediction by the EG model) while $M(r_i)$ and $\sigma(r_i)$ are provided by the weak lensing data. The relevant quantity to compare the model to the data is a reduced $\chi^2$, which is calculated as $\chi^2_{d.o.f} = \chi^2 / N_{d.o.f}$, where $N_{d.o.f.}$ is the number of degrees of freedom. 

As shown previously, the best qualitative agreement is the radial region around the virial radius.
In what follows, we measure each of the cluster mass profiles with a step $0.1R_{200}$ and for example in the range from $0.2R_{200}$ to $R_{200}$ that gives us $8$ data points per clusters and $184$ data points in total as we have 23 clusters in our data sample. The total $N_{d.o.f.} = 181$ since the Einasto matter density model has three free parameters. 

In spite of the fact that at $\sim R_{200}$ the predicted by the EG model the apparent dark matter is similar to the observed dark matter, quantitatively we find that the profiles predicted by EG differ from the observed profiles by $> 5\sigma$. The best agreement we find is within the narrow range $0.55R_{200} \leqslant r \leqslant 0.75R_{200}$, where the EG model is only ruled out at the $2\sigma$ level.

\begin{figure}[t]
\centering
\includegraphics[width=1\linewidth]{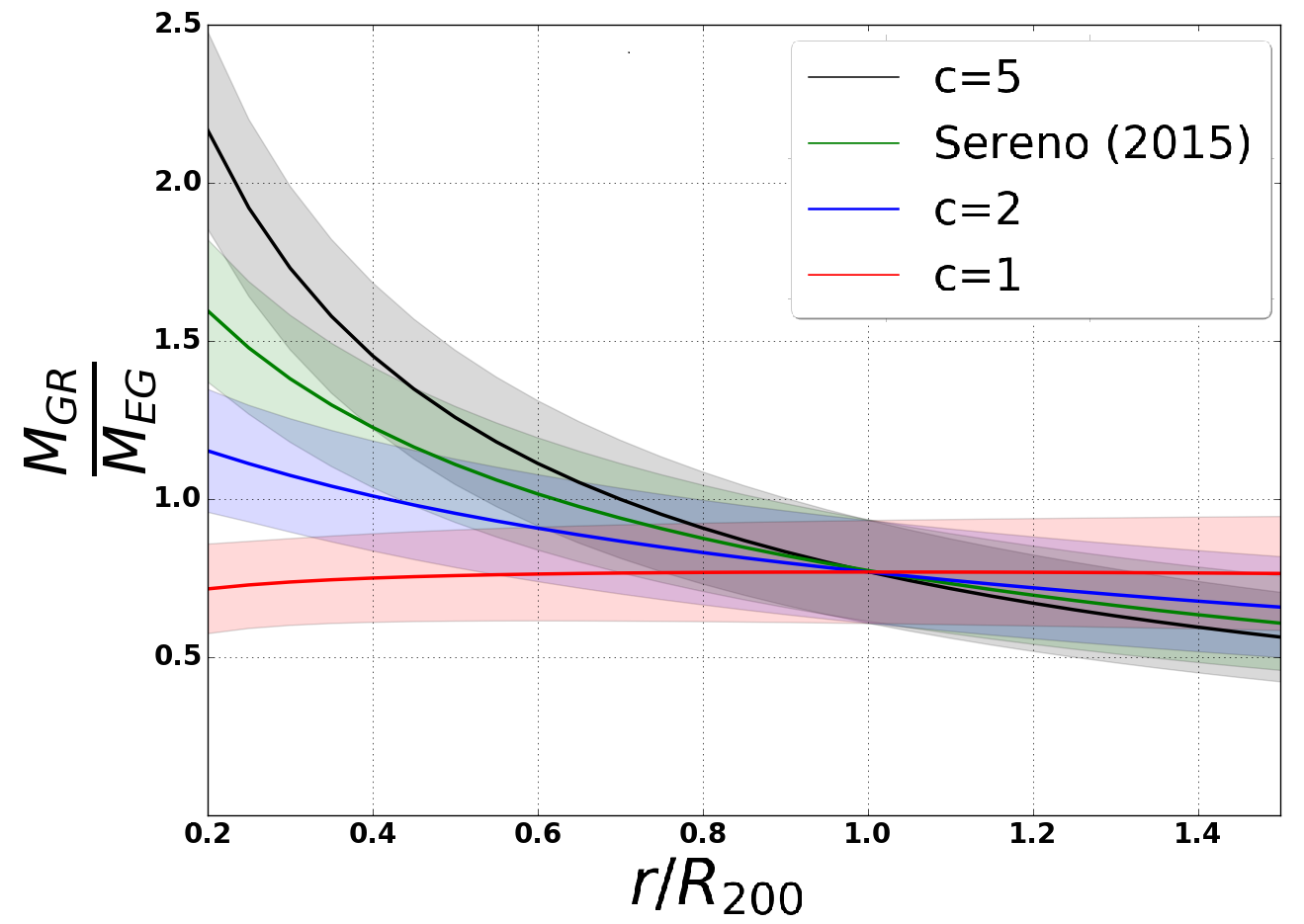}
\caption{The mass ratio $\frac{M_{GR}}{M_{EG}}$ of the observed dark matter ($M_{GR}$) to the predicted by the EG model apparent dark matter ($M_{EG}$) at different concentrations $c_{200}$ (\ref{conc_sereno}). Solid lines and shaded regions around them are the means and $68.3\%$ error bars around the means. Green color corresponds to the case with the concentrations $c_{200}$ that are given by \citet{Sereno:2014aea}. Red, blue and black colors correspond to the concentrations $c_{200} = 1, 2$ and $5$ with $M_{200}$ given by \citet{Sereno:2014aea}. As it was pointed out in subsection \ref{weak_lensing_data}, the mean concentration of the data from \citet{Sereno:2014aea} is $<c_{200}> = 3.15$. It can be seen from the plot that the EG model prefers smaller concentrations.}
\label{plot:c_new}
\end{figure}

Having uncertainties of the baryon density profiles could not easing significantly the level of the precision of the constraint of the EG model. To confirm this statement we add some error of the baryon profiles by treating $\sigma(r_i)^2$ in the formula (\ref{chi_fla}) as a sum of the squares of the errors of the weak lensing ($\sigma_{weak}$) and baryon masses ($\sigma_{bar}$), i.e. $\sigma(r_i)^2 = \sigma_{weak}(r_i)^2 + \sigma_{bar}(r_i)^2$. Placing uncertainties on the baryon matter even half of the uncertainties of the weak lensing data [i.e. $\sigma_{bar}(r_i) = 0.5\sigma_{weak}(r_i)$] does not decrease significantly the level of constraining EG model in the range $0.3R_{200} \leqslant r \leqslant R_{200}$ as it is still $\sim 5\sigma$. However, with these baryon matter uncertainties the EG model is compatible with the observations at almost $1\sigma$ level in the "narrow" range. 

Given that the amplitude predicted by EG is reasonably well represented by the model, we focus our comparison on the profile shapes. Fig. \ref{mass_ratio_23} shows the mass ratio $\frac{M_{GR}}{M_{EG}}$ of the observed dark matter ($M_{GR}$) and the predicted by the EG model apparent dark matter ($M_{EG}$). One can see that the observed dark matter is almost two times higher than the apparent dark matter in the area close to the cores ($0.1R_{200}$) of the galaxy clusters (around 40\% higher at $0.3R_{200}$) and it can also be seen that the mass profiles of the dark matter and the apparent dark matter are very different. EG  underestimates the dark matter mass in the regions closer to the core while overestimating the mass in the regions beyond approximately $0.9R_{200}$. At the current stage we must claim that the EG model is unable to describe the real observational data at Mpc scales.

\subsection{Systematic uncertainty from concentration}\label{playing_with_c}

\begin{figure}[t]
\centering
\includegraphics[width=1\linewidth]{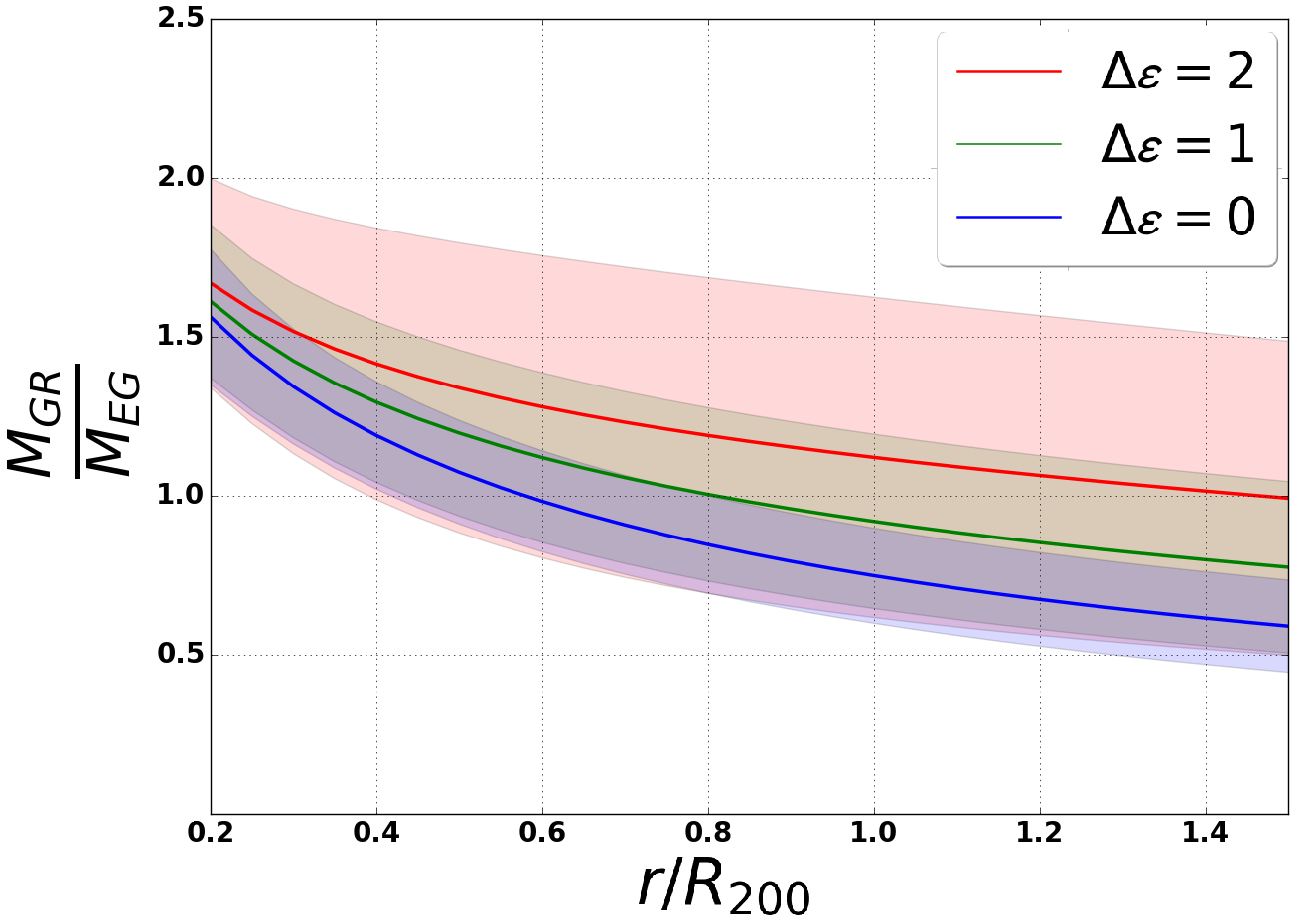}
\caption{The mass ratio $\frac{M_{GR}}{M_{EG}}$ of the observed dark matter ($M_{GR}$) and the predicted by the EG model apparent dark matter ($M_{EG}$) at different values of steepness parameter $\epsilon$ (described in the formula \ref{beta_profile}). Solid lines and shaded regions around them are the means and $68.3\%$ error bars around the means. Baryon matter distribution in our sample have rather small steepness: $<\epsilon> = 1.69$ for 20 clusters and zero $\epsilon$ for the three clusters with double beta profiles (\ref{double_beta}). However, in general steepness parameter is higher (for example it is $<\epsilon> = 3.24$ in \citet{Vikhlinin:2005mp}). To take that into account we have increased $\epsilon$ of the 20 clusters by $1$ (green) and by $2$ (red) and that made steepness parameter to be $<\epsilon> = 2.69$ and $<\epsilon> = 3.69$ respectively. Blue color corresponds to the implementation of the data with the original steepness parameters.   
}
\label{plot:epsilon_new}
\end{figure}

As it was discussed above (see subsection \ref{weak_lensing_data}), the mass-concentration relation of the galaxy clusters is a source of systematic uncertainty. We can include these systematics in the following way: $\sigma(r_i)$ in the formula (\ref{chi_fla}) is now a sum of statistical and systematical uncertainties, i.e. $\sigma(r_i)^2 = \sigma_{weak}(r_i)^2 + \sigma_{sys}(r_i)^2$. We neglect $\sigma_{bar}(r_i)$ here as discussion of the baryon uncertainty was done in the previous subsection.
We define $\sigma_{sys}(r_i)$ as the difference between the true value of the $M_{DM, true}$ [i.e. at the concentration which is given by the data (\ref{conc_sereno})] and $M_{DM, new}$ (at the concentration motivated by \citet{Groener:2015cxa}),
\begin{equation}
    \sigma_{sys}(r_i) = M_{DM, true} - M_{DM, new}.
\end{equation}
Through this technique, we allow the systematic uncertainty in the concentration to impact the uncertainty on the amplitude of the profiles, but not the shape.
We consider the effect of systematic uncertainties by concentrations up to $c_{200, new} = 10$.  We focus our analysis only on the range ($0.3R_{200} \leqslant r \leqslant R_{200}$) where the mass densities are measured with the step $0.1R_{200}$.
The effect of the systematic uncertainty starts to be noticeable at $c_{200, new} \approx 4.1$ were the median $\sigma(r_i) / \sigma_{sys}(r_i) \approx 5$. This effect pushes the constraint level down to $\sim 3\sigma$ and at $c_{200, new} = 10$ the EG model is compatible with the observations at $1\sigma$.

\subsection{Systematic shape bias from concentration}\label{playing_with_c_bias}

\begin{figure}[t] 
\centering
\includegraphics[width=1.03\linewidth]{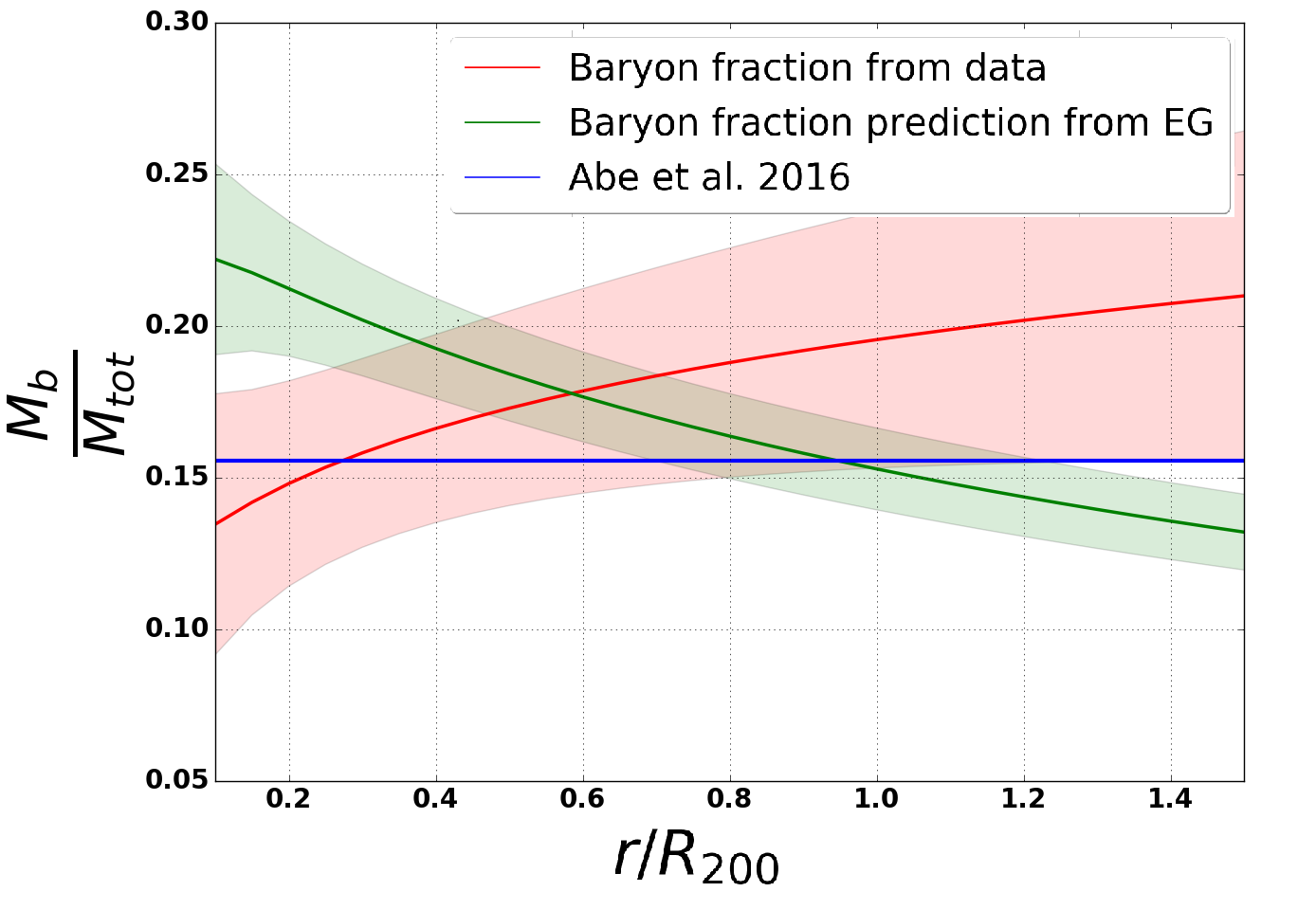}
\caption{The ratio of baryon mass to the total mass of the galaxy cluster as a function of radius of the observed data set of 23 galaxy clusters. Red line and red shaded region represent the baryon fraction of the observed clusters, i.e. $M_b / M_{tot, GR}$, where $M_b$ is the observed baryon mass, $M_{tot, GR}$ is the total mass from the weak lensing data and this result correlates with other results \citep{2009ApJ...703..982G, 2010MNRAS.407..263A} as we expect to see higher baryon fraction for heavier galaxy clusters and the average mass of the clusters in our sample is high ($<M_{200} = 1.14 \times 10^{15} M_{\odot}$). Green line and green shaded region correspond to the effective baryon fraction which is predicted by the EG model, i.e. $M_b / M_{tot, EG}$, where $M_{tot, EG}$ is the total mass predicted by the EG model, i.e. the sum of the apparent dark matter and the baryon matter. Solid lines are the mean values and shaded regions are $68.3\%$ error bars around the means. One can observe that the EG model prediction diverge from the observed baryon fraction starting from the cores of the clusters up to  $\sim 0.6 R_{200}$ which means that the EG model predicts that the baryon fraction is the biggest in the regions around the core of the clusters while the observations predict the baryon fraction to increase with a distance from the core. Interestingly, the baryon fraction prediction of the EG model agrees well at around $R_{200}$ with the baryon fraction from the CMB \citep{Ade:2015xua} (see blue flat line).
}
\label{baryon_fraction}
\end{figure}

\begin{figure*}[t]
\centering
\includegraphics[width=0.49\linewidth]{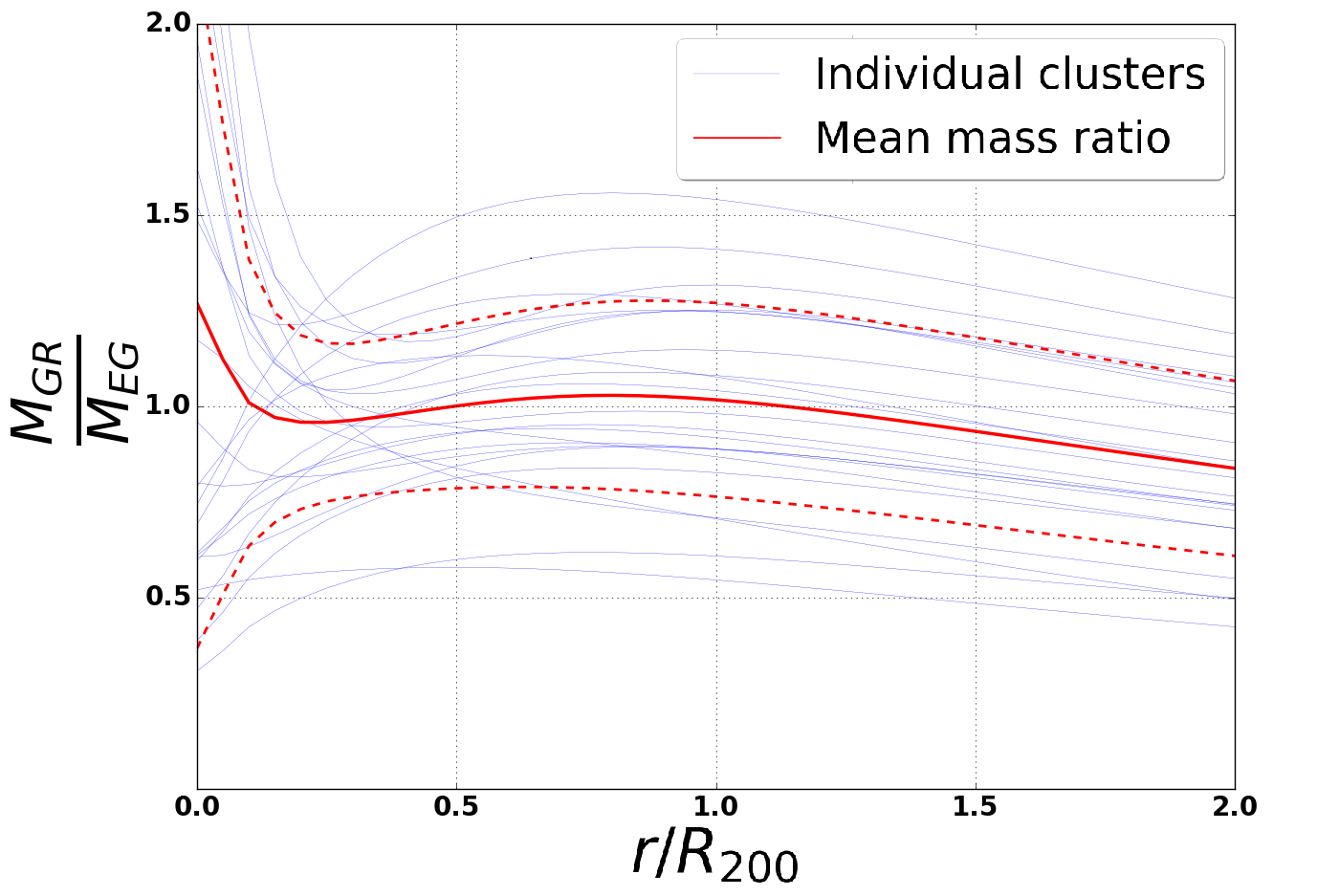}
\includegraphics[width=0.49\linewidth]{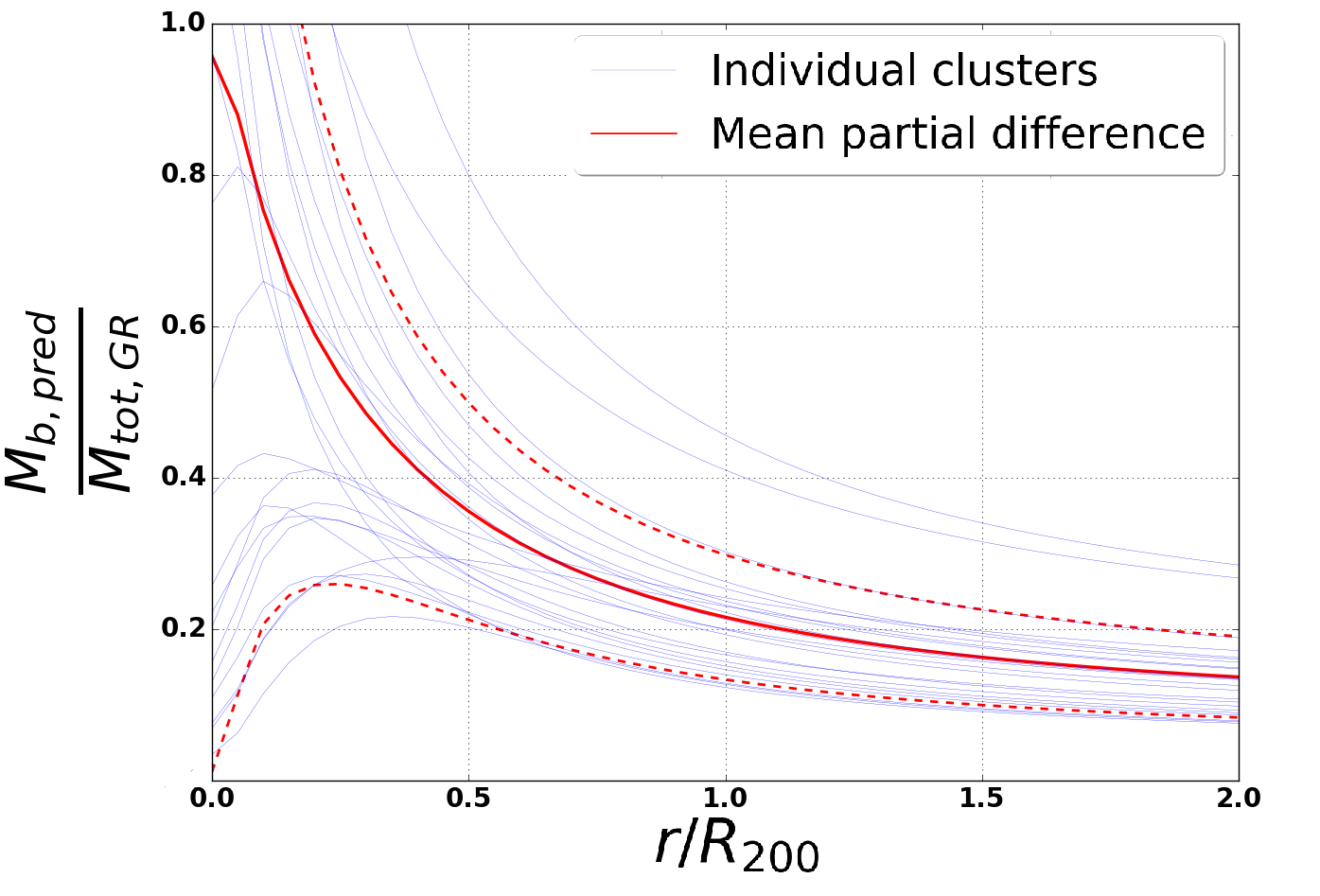}
\caption{Left: the predicted dark matter mass ratio  $M_{GR} / M_{EG}$ in the case of the baryon fraction  $M_{b, pred} / M_{tot, GR}$ in the form from the right figure. $M_{b, pred}$ is the predicted baryon matter, $M_{tot, GR}$ is the total observed mass from the weak lensing data,  $M_{GR}$ is the observed dark matter and  $M_{EG}$ is the predicted apparent dark matter with the predicted baryon matter $M_{b, pred}$. For the EG model to be able to properly describe the weak lensing data (left figure) the baryon fraction should have rather weird shape (right figure). One of the biggest problems with such baryon fraction is the huge amount of baryon matter in the core (i.e. baryon fraction is close to unity) which is in total contradiction with the observations (compare with red line on Fig. \ref{baryon_fraction}). }
\label{baryon_predict}
\end{figure*}

An alternative approach to evaluate systematic uncertainty due to the mass-concentration relation is to fix the mass measurements with our current errors while allowing the profile shapes to be more uncertain. As we can see from Fig. \ref{plot:c_new}, if we assume that the cluster weak-lensing inferred masses are unbiased, the EG model becomes more consistent with the data at $c_{200} \approx 2$. While small, this average value for the NFW concentration of the weak-lensing mass profiles of massive clusters is close to those obtained in simulations \citep{Groener:2015cxa, Klypin:2014kpa, Correa:2015dva}.

\subsection{Baryon profile bias}\label{steepness}
Three clusters from \citet{Giacintucci:2017xyd} utilize double beta profile (\ref{double_beta}) which does not take into account steepness parameter $\epsilon$ in Eq. (\ref{beta_profile}). The remaining 20 clusters in our sample have average steepness parameter $<\epsilon> = 1.69$ which is significantly smaller than the average steepness parameter $<\epsilon_{V}> = 3.24$ of \citet{Vikhlinin:2005mp} dataset. Increasing $\epsilon$ in our data rotates the apparent DM distribution curve and shifts it upwards which makes the EG prediction of the apparent DM more consistent with the observation of DM (see Fig. \ref{plot:epsilon_new}). Recent results from \citet{Ettori09,Eckert12} suggest that the baryon profiles are in fact much steeper than the original beta profile and in agreement with the high $\epsilon$ values from \citet{Vikhlinin:2005mp}.

\subsection{Other systematics}
One of the assumptions of the EG model, which was discussed above in the introduction, is the fixed value of the Hubble parameter. To test this assumption we divided by redshifts our data sample of 23 galaxy clusters into two bins, i.e. one bin contained 11 clusters with the lowest redshifts ($<z> = 0.17$) and the second bin contained 12 clusters with the highest redshifts ($<z> = 0.25$). Analysis of both bins produced almost completely identical results and that validates the fixed Hubble parameter assumption.

The second assumption which we made on the data is that the hot gas represents the total baryon mass of the clusters which is not totally true as stars contribute as well. However, stellar mass is less than $10\%$ \citep{2009ApJ...703..982G, 2010MNRAS.407..263A, 2013A&A...555A..66L} of the hot gas for the clusters with the masses we use in this paper ($<M_{200}> = 1.14 \times 10^{15} M_{\odot}$). To check this assumption, we increased the baryon mass by $10\%$ which shifted the mass ratio $\frac{M_{GR}}{M_{EG}}$ in Fig. \ref{mass_ratio_23} only by approximately $0.05-0.08$ or changed this ratio by around 6\%. This small shift  in the mass ratio not only does not change the precision of constraining the EG model, but also does not change at all the main conclusion of incompatibility of the EG model with the galaxy clusters. So, the assumption of neglecting stellar masses is totally valid.

\section{Discussion}
\label{!discussion}

In this section, we discuss the consequences of the current EG predictions in the context of the observational data. We also explore alternatives to our fiducial analysis which could bring the EG predictions and the data into better agreement.

\subsection{Effect on the baryon fraction}
One of the consequences of the EG model is in the distribution of the baryons in clusters.
We can define the effective baryon fraction which is predicted by the EG model by introducing the following ratio 
\begin{equation}\label{fb_eg}
    f_{b, EG} = \frac{M_b}{M_{tot, EG}},
\end{equation}
where $M_b$ is the observed baryon mass and $M_{tot, EG}$ is the total mass which is predicted by the EG model.

The results of Fig. \ref{baryon_fraction} imply that the EG effective baryon fraction is different in many aspects from the observed baryon fraction with the total mass $M_{tot, GR}$ defined by the weak lensing data. The first difference is the shape of the lines in Fig. \ref{baryon_fraction}: the EG model has a monotonically decreasing behavior while the data shows that the baryon fraction is an increasing with the radius function. In agreement with \citet{Nieuwenhuizen:2016uxv} this means that the EG predicts baryons to be concentrated in the region around the cores of the galaxy clusters while the observations imply that the baryons are actually spread in the broader regions with highest fraction in the outskirts of the clusters. Second, the effective baryon fraction is almost twice as high close to the core (at $r \approx 0.1R_{200}$) which should be detected as it implies brighter cluster cores than we would observe in GR. This effect could be actually smaller if BCGs would be correctly taken into account by weak lensing data. In spite of these differences, the EG model predicts correctly the baryon fraction at the distances approximately $0.4R_{200} \leqslant r \leqslant 0.8R_{200}$. Additionally, the EG model predicts the effective baryon fraction to be close to $15.6\%$ (the number which is expected from the CMB observations \citep{Ade:2015xua}) at the distances close to $R_{200}$.

One of the tenets of EG is that there is no particlelike dark matter. In the case of a flat universe, the only two contributions to the energy density are baryons and dark energy \citep{Ade:2015xua}. We can build a toy model for how the baryons should be distributed in EG such that at the core of a virialized system one finds $\sim 100\%)$ of the baryons while in the outskirts the EG baryon fraction falls to the global value of 5\%-10\%. This toy model is shown in Fig. \ref{baryon_predict} right. If this toy model describes how the real baryons are distributed in our Universe we would find a high level of consistency between the dark matter profiles from observed weak lensing data and what EG predicts for the apparent dark matter (see Figure \ref{baryon_predict} left). This is just a toy model, but it is an example of how one could achieve closer agreement between the EG predictions and the current observations.

\subsection{Modifying EG}\label{modif_eg}

As opposed to reconsidering the distribution of the baryons inside clusters, one could alter the maximal strain of the EG model as described in Sec. \ref{!theory} in Eq. (\ref{inequality}). Recall that we chose equality in the inequality of the EG model in Eq. (\ref{inequality}). We could have chosen some form away from its maximum value.  As a new toy model, we propose a modification to the EG model which consists in changing $r'^2 \rightarrow r_0 r'$ in the denominator of the r.h.s. of the Eq. (\ref{Main_eg}). For $r_0 = 1.2$Mpc, the l.h.s. is smaller than its maximum value until beyond this radius. In the case $r_0 = 1.2$ Mpc the result is consistent with the observations (see Fig. \ref{mass_ratio_modif_eg}). While the modification is based purely on phenomenological ground it might help in developing the theory of the EG model as we can see that the data favor the proposed form instead of the original form (\ref{Main_eg}). This result leads to the conclusion that while by default equality is chosen in most of the works related to the testing and development of the EG theory, it is not necessarily the right or only choice.

\subsection{Combining systematics}
\begin{figure}[t] 
\centering
\includegraphics[width=1.03\linewidth]{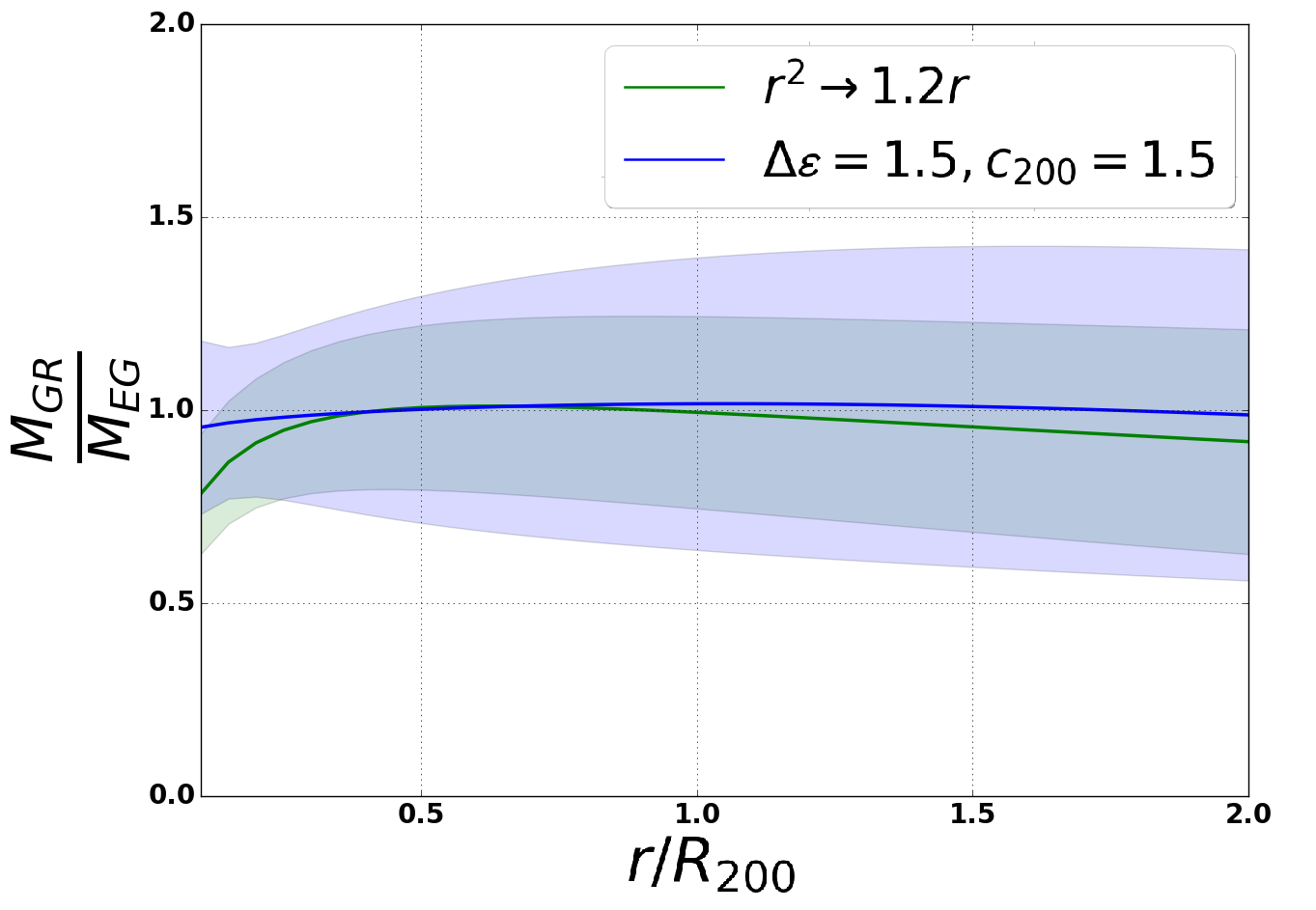}
\caption{The mass ratio $\frac{M_{GR}}{M_{EG}}$ of the observed dark matter ($M_{GR}$) to the apparent dark matter ($M_{EG}$). Solid lines and shaded regions are the means and $68.3\%$ error bars around the means. Green color corresponds to the phenomenological modification of EG prediction (see subsection \ref{modif_eg}) in the case of substituting $r^2$ in the denominator of the r.h.s. of the Eq. (\ref{Main_eg}) by $1.2r$. Blue color corresponds to the adjusting both weak lensing data (shifting concentration parameter so it is $c_{200} = 1.5$ for all the data (see subsection \ref{playing_with_c} for motivation of this modification)) and baryon matter distribution (increasing steepness parameter by $\Delta\epsilon = 1.5$ for all the clusters (see subsection \ref{steepness} for motivation of this modification)). It can be seen that both modifications presented in the figure make EG model to be consistent with the observed data as the mass ratio $\frac{M_{GR}}{M_{EG}} \approx 1$ in the radial region $0.3 \leq r/R_{200} \leq 2$.
}
\label{mass_ratio_modif_eg}
\end{figure}

As it was mentioned in the section \ref{!test_eg}, concentration parameter ($c_{200}$) of the weak lensing and the steepness parameter ($\epsilon$) could be changed to make EG be more compatible with the observed data. Moreover, by adjusting both of these parameters at the same time the prediction of the EG model correlates nicely with the observed data (see Fig. \ref{mass_ratio_modif_eg}).

\section{Conclusions}
\label{!concl}

The first attempt to test emergent gravity was done by \citet{Nieuwenhuizen:2016uxv}, where in contrast to our approach of using only weak lensing in determining matter profiles, combination of strong and weak lensing data (which compliment each other and overall better than weak lensing along determine matter profiles \citep{Umetsu:2013dn}) of one cluster A1689 showed that EG does not work in the region up to $0.4-0.5 R_{200}$, while inclusion of neutrinos into EG framework helps to achieve a very good fit. \citet{Brouwer:2016dvq} showed that the EG model is in good agreement with the galaxy data. \citet{Ettori:2018tus} tested the EG theory with 13 clusters in the narrow small redshifts range ($z \approx 0.047-0.091$) with reconstructed hydrostatic mass profiles which have non-negligible hydrostatic bias due to nonthermal pressure sources. By analyzing 4 clusters, \citet{ZuHone:2019hdt} confirmed conclusion of current manuscript as well as supported results of \citet{Nieuwenhuizen:2016uxv} that at small radii ($\sim 3-100$ kpc), EG produces a bad fit to the data. 

In this work, the cluster data set was extended and resulted in utilization of 23 galaxy clusters in wider radial ($0.1R_{200}-2R_{200}$) and redshift ($0.077-0.289$) ranges. In addition to testing the nominal EG model, we consider an extension to the basic predictions of the framework (see also \citet{Hossenfelder:2017eoh}).

EG provides good results only in the area near the virial radius and by taking into account the cores and the outskirts, the mass profile shape differences allow us to rule out EG at $>5\sigma$. However, given our current level of systematic errors in the observed shape profiles, our results lead to the conclusion that the EG model is a viable alternative to dark matter in the range $0.3 \le r \le 1$R$_{200}$. Under the nominal assumptions (i.e., without systematics), EG favors a radially decreasing baryon fraction which peaks in the cluster core (this effect could be slightly amplified due to the BCG not always taking into account by weak lensing data). This is a different baryon fraction profile when compared the standard dark matter model (see \citet{Ade:2015xua}). 

The EG model predicts a flatter shape of the dark matter mass distribution than the observed data, as well as steep x-ray gas density profiles. One of the successes of the model is that the observed weak lensing data and the predicted apparent dark matter are almost identical in the region close to $R_{200}$. 

Finally, we investigate the level of systematic errors needed to reach good agreement between EG and the data. We find that within the current systematic limits, there are combinations of shape profiles which can match EG to the data. Likewise, we investigate whether the EG model itself has the flexibility to better match the data and we find that it does through a lowering of the maximal strain.  Given the level of systematic uncertainties in the data, as well as the depth of the theoretical framework, we are unable to formally rule out in the wide region (i.e. $0.3 \le r \le 1$R$_{200}$) the EG model as an alternative to dark matter in galaxy clusters.

\section{Acknowledgments}
This material is based upon work supported by the National Science Foundation under grant no. 181273. This research has made use of the VizieR catalogue access tool, CDS, Strasbourg, France.

\bibliographystyle{apsrev4-1}
\bibliography{main}

\end{document}